July19, 2007

# The quantummechanical wave equations from a relativistic viewpoint

Engel Roza[1]

**Summary**

A derivation is presented of the quantummechanical wave equations based upon the Equity Principle of Einstein's General Relativity Theory. This is believed to be more generic than the common derivations based upon Einstein's energy relationship for moving particles. It is shown that Schrödinger's Equation, if properly formulated, is relativisticly covariant. This makes the critisized Klein-Gordon Equation for spinless massparticles obsolete. Therefore Dirac's Equation is presented from a different viewpoint and it is shown that the relativistic covariance of Schrödinger's Equation gives a natural explanation for the dual energy outcome of Dirac's derivation and for the nature of antiparticles. The propagation of wave functions in an energy field is studied in terms of propagation along geodesic lines in curved space-time, resulting in an equivalent formulation as with Feynman's path integral. It is shown that Maxwell's wave equation fits in the developed framework as the massless limit of moving particles. Finally the physical appearance of electrons is discussed including a quantitative calculation of the jitter phenomenon of a free moving electron**.**

**Introduction**

Schrödinger's equation, being the first and most prominent quantummechanical wave equation, has historically been derived in a rather heuristic way [1]. To provide a theoretical basis and relativistic versions of it, Einstein's energy relationship for moving particles is applied in combination with some basic hypotheses, such as rules to transform momenta into operators on wave functions [2]. These attempts resulted into the Klein-Gordon equation, the validity of it of which still gives rise to debates, as the attached probability density function, which determines to find a particle at certain time at a certain position, is not positive definite [3]. To overcome this difficulty, Dirac proposed, by a brilliant mathematical manipulation, a solution which did not show this difficulty [4]. His Dirac's equation revealed the basics of an already presumed property of particles, called spin, which explained up to then uncomprehensible behaviors of the electron. The non-relativistic limit of this equation corresponds with a heuristic formulation given before by Pauli [5]. For spinless particles however, the Klein-Gordon equation, in spite of the critics, is still considered as being applicable. Much later, independently of each other, Tomanaga, Schwinger and Feynman elaborated a fundamental extension of wavemechanics theory, which formulated interaction of particles of different nature, such as the interaction of electrons and photons [6]. Feynman thereby introduced an enlightening alternative view on wavemechanics, known as the path integral method [7].

It is the purpose of this paper to provide a simple and consistent view on the wave equations, which will be shown to result in some divergent views as compared with today's perception. This holds in particular for the relativistic version of Schrödinger's equation for spinless particles. In addition it will be shown that Dirac's equation, although originally conceived in an attempt to produce a relativistic version of Schrödinger's equation, does not necessarily require a relativistic starting point, but merely the hypothesis of an additional energy dimension. This view makes it possible to formulate a relativistic and a non-relativistic version of it. Yet another result of the views developed in this paper is a simple understanding of Feynman's formulation of the path integral.

---

1. Engel Roza is an electrical engineer, retired from Philips Research Labs., Eindhoven, The Netherlands.

It is very well possible that the contents of the paper does not contain anything new for theoretical physicists. If so, the author, who is not an expert but just a scientist interested in physics, was unable to find it in literature accessible to him. So, he hopes that the paper at least is elucidating for non-experts with a similar interest.

Next to the temporal dimension, the derivation of the various wave equations will be based upon a consideration in a single spatial dimension. It is believed that the generalisation towards three spatial dimensions will be not a basic problem. The views developed in this paper will be based on the principles of General Relativity, in particular on the application of Einstein's principle of equity. This makes it different from derivations from Einstein's energy relationship of moving particles, the scope of which is restricted to Special Relativity. So, it is believed that our starting point is a more general one than commonly adopted in theoretical explanations of the basic wave equations.

## Basics

Wavemechanics reflect a statistical view on physics. It means that assigning wavefunctions to moving particles does not mean that a single harmonic function is assigned to an individual moving particle but instead that a ensemble of harmonic functions is assigned to an ensemble of moving particles. Because of the underlying assumption of linearity of the theory we may assign to a single particle a collection of harmonic wave functions. Although a single member of this collection is non-causal, the total of all wave functions in this collection, if suitably composed, may guarantee causality and locality of the particle if, according to the Born interpretation, the energy of the total wave function is seen as the probabiliy density of finding the particle at a certain time at a certain position. Moreover, if care is given to the statistical nature of the theory, an individual wave function can be conceived as the wave function of an elementary *pseudo*-particle in spite of the non-causal character of this individual wave function. This conception makes it possible to transform a motion equation of a *real* particle into a wave function by suitable rules. These rules form the basic hypothesis for the development of a mathematical theory of wave mechanics. This all, of course is known basics, but nevertheless important as we shall challenge in the next sections the rules.

By this we have stated the importance of the motion equation of a single particle. To be generic, the motion equation should be formulated in a relativistic format. As the particle may move in an energy field, the formulation should be made in terms of General Relativity. Special Relativity is not enough, as Special Relativity presupposes a constant speed. So let us see where General Relativity brings us.

The generic relativistic motion equation is given by Einstein's geodesic equation. In fact this is not a single equation but a set of equations, i.e. an equation for each dimension, one temporal and three spatial ones. Written in the general compact formulation, the interpretation requires some study. However, if we consider next to the temporal dimension a single spatial one, the equation becomes already more comprehensible. We may write it into the following form [8]:

$$\frac{d^2 x}{d\tau^2} + \frac{1}{2g_{xx}}\left[\frac{\partial g_{xx}}{\partial x}\left(\frac{dx}{d\tau}\right)^2 - \frac{\partial g_{tt}}{\partial x}\left(\frac{dt'}{d\tau}\right)^2 + 2\frac{\partial g_{xx}}{\partial t'}\left(\frac{dx}{d\tau}\frac{dt'}{d\tau}\right)\right] = 0$$

$$\frac{d^2 t'}{d\tau^2} + \frac{1}{2g_{tt}}\left[\frac{\partial g_{t't'}}{\partial t'}\left(\frac{dt'}{d\tau}\right)^2 - \frac{\partial g_{xx}}{\partial t'}\left(\frac{dx}{d\tau}\right)^2 + 2\frac{\partial g_{tt}}{\partial x}\left(\frac{dx}{d\tau}\frac{dt'}{d\tau}\right)\right] = 0 \; . \qquad (1,a,b)$$

What does it mean? We see the following parameters: $x$, $\tau$, $t'$, $g_{xx}$ and $g_{tt}$. Of course $x$ is the spatial coördinate. Furthermore we have two different time coördinates. The parameter $t'$ is the normalized time for the stationary observer, i.e. $t' = jct$, wherein $c$ is the light velocity and $j = \sqrt{-1}$. The parameter $\tau$ is proper time, i.e. the wrist time of a co-moving observer (co-moving with the particle). The parameters $g_{xx}$ and $g_{tt}$ are elements of the so called metric tensor. They determine the way how the frame of coördinates $\xi, \tau'$ of the co-moving observer is transformed (by considering $\xi(x, t')$ and $\tau'(x, t')$ )into the frame of coördinates $(x, t')$ of the stationary observer. In particular:

$$\left(\frac{\partial \xi}{\partial x}\right)^2 + \left(\frac{\partial \tau'}{\partial x}\right)^2 = g_{xx} \quad \text{and} \quad \left(\frac{\partial \xi}{\partial t'}\right)^2 + \left(\frac{\partial \tau'}{\partial t'}\right)^2 = g_{tt}, \text{with } \tau' = jc\tau. \tag{2,a,b}$$

The geodesic equation is a result of Einstein's principle of equity which states that the forced movement of the particle seen by the stationary observer is seen as a free movement seen by the co-moving observer. Therefore the geodesic equation is nothing more than a transformation of the free movement equation by merely transforming the coördinates.

Note that the time parameter of the geodesic equation is the proper time. Note also that temporal and spatial parameters occur on par, i.e. both equations are fully symmetrical.

The parity is lost after introduction a basic step in further evaluation. This step is the assumption of stationarity. This means that we shall assume that the elements of the metric tensor are independent of time. This has to do with the nature of the energy field in which the particle is supposed to move. In many cases, such as movement in a gravity field or in a static electromagnetic field, this assumption is justified. If however this static charachter is violated, for instance as a consequence of particular forms of particle interaction, a reconsideration will be necessary. Such a reconsideration is beyond the scope of this paper.

Because of the stationary condition the geodesic equation simplifies to:

$$\frac{d^2 x}{d\tau^2} + \frac{1}{2g_{xx}} \left[\frac{dg_{xx}}{dx}\left(\frac{dx}{d\tau}\right)^2 - \frac{dg_{tt}}{dx}\left(\frac{dt'}{d\tau}\right)^2\right] = 0$$

and

$$\frac{d^2 t'}{d\tau^2} + \frac{1}{g_{tt}}\frac{dg_{tt}}{dx}\left(\frac{dx}{d\tau}\frac{dt'}{d\tau}\right) = 0. \tag{3,a,b}$$

The latter equation can be easily integrated, resulting in a linear differential equation:

$$\frac{dt'}{d\tau} = \frac{k_0}{g_{tt}}, \text{ wherein } k_0 \text{ is an integration constant to be determined.} \tag{4}$$

In addition we may consider a redundant equation which makes further use of (3a) obsolete. This equation is the formulation of the invariance of the local space-time interval, i.e.:

$$d\tau'^2 = g_{xx}dx^2 + g_{tt}dt'^2 \quad \text{so that:} \quad \left(\frac{d\tau'}{d\tau'}\right)^2 = g_{xx}\left(\frac{dx}{d\tau'}\right)^2 + g_{tt}\left(\frac{dt'}{d\tau'}\right)^2, \tag{5}$$

which is equivalent with

$$g_{xx}\left(\frac{dx}{d\tau}\right)^2 + g_{tt}\left(\frac{dt'}{d\tau}\right)^2 = -c^2. \tag{6}$$

Applying (6) to (4) we get:

$$\left(\frac{dx}{d\tau}\right)^2 = \frac{1}{g_{xx}}\left[-c^2 - \frac{k_0^2}{g_{tt}}\right]. \tag{7}$$

It can be shown that elaboration on the basis of (3a) instead of (6) gives the same result. The prime importance of (3,a,b) is the split into two equations.

Equations (4) and (7) together form an excellent set to derive wave equations for a single spatial dimension by basic rules in which momenta are transformed into operators on wave functions. Because General Relativity is fully taken into account there are reasons to believe that this set provides a more fundamental basis than just Einstein's energy relationship. In the subsequent sections we shall show how to transform this relativistic basic motion equation into wave equations.

## Schrödinger's equation

If we formulate (7) and (4) in momentum notation we have:

$$p_0 = \frac{k_0}{g_{tt}} \quad \text{and} \quad p_x^2 = \frac{1}{g_{xx}}\left[-c^2 - \frac{k_0^2}{g_{tt}}\right], \quad \text{wherein} \quad p_0 = \frac{dt'}{d\tau} \quad \text{and} \quad p_x = \frac{dx}{d\tau}. \tag{8,a,b}$$

Note that $p_0$ and $p_x$ actually are momenta per unit of restmass, so that, strictly spoken, a proportionality factor with numerical value 1 is required to correct for the dimensionality.

From (8b) it follows that:

$$p_x = \pm \sqrt{\frac{1}{g_{xx}}\left[-c^2 - \frac{k_0^2}{g_{tt}}\right]}. \tag{9}$$

The next step is the adoption of basic rules whereby momenta are transformed into operators on wave functions. Note that we have to do with linear equations only, albeit that the spatial equations have been naturally splitted up into two separate equations. Anyhow, the ambiguity as in the derivation of the Klein-Gordon equation does not show up. And, as stated by Dirac, linearity in the temporal differential quotient is a prerequisite to guarantee positive definiteness of the resulting probability density function, which is a shortcoming of the Klein-Gordon equation.

The rules are defined as: $\hat{p}_i = \frac{\tilde{h}}{j}\frac{\partial}{\partial x_i}$ wherein $x_1 = x$ and $x_0 = t'$.

Note: We propose here a full relativistic version of the transform rules, derived from proper time momenta with strict parity in temporal domain and spatial domain. Usually a less general format is used, based upon stationary time momenta and Hamiltonian [9]. The author believes that this gives a loss in maintaining the relativistic character of derivations.

From (8a) we have:

$$\frac{\tilde{h}}{j}\frac{\partial}{\partial t'}\Psi(x,t') = \frac{k_0}{g_{tt}}\Psi(x,t')$$

and from (8b):
$$\frac{\tilde{h}}{j}\frac{\partial}{\partial x}\Psi(x,t') = \pm\sqrt{\frac{1}{g_{xx}}\left[-c^2 - \frac{k_0^2}{g_{tt}}\right]}\Psi(x,t'). \tag{10,a,b}$$

These equations are not easily simultaneously solvable. However they can if separation of variables is possible. As we will see this is only the case if $g_{tt}$ is constant. An obvious case for which this is true is the case that the particle moves at a constant speed. Of course we are interested in other cases as well, but let us first consider this case of a free moving particle. So, we state that:

$$\Psi(x,t') = \Psi_x(x)\Psi_t(t).$$

From (10a) we get under consideration that $t' = jct$:

$$\frac{\tilde{h}}{c}\frac{\partial \Psi_t}{\partial t} = \frac{k_0}{g_{tt}}\Psi_t$$

$$\frac{\tilde{h}}{j}\frac{\partial \Psi_x}{\partial x} = \pm\sqrt{\frac{1}{g_{xx}}\left[-c^2 - \frac{k_0^2}{g_{tt}}\right]}\Psi_x$$

and from (10b): $\tag{11a,b}$

So we see that $\Psi_t$ depends only on $t$ exclusively if $g_{tt}$ is a constant. This means that the clock dilatation in the frame of the co-moving observer is constant, implying a constant speed of the particle. Under this condition the solution of (11a) is:

$$\Psi_t = k_1 \exp\left[-j\frac{E}{\tilde{h}}t\right] \text{ wherein } E = \frac{1}{j}\frac{ck_0}{g_{tt}}. \tag{12}$$

As the particle is supposed to move at a constant speed we may invoke the properties of the Lorentz transform to give an interpretation of the constant $E$. In that case we have $g_{tt} = g_{xx} = 1$ and:

$$\frac{dt}{d\tau} = \frac{1}{w} \text{ with } w^2 = 1 - \frac{v^2}{c^2} \text{ and } v_x = \frac{dx}{dt}. \tag{13}$$

Applying these conditions to (4) we find: $\quad k_0 = j\frac{c}{w}. \tag{14}$

Substituting this expression into (12) gives:

$$E = \frac{c^2}{w}. \tag{15a}$$

This expression can be interpreted as the particle's relativistic energy per unit of mass. This is clear if we apply the usual expansion:

$$E = c^2 + \frac{1}{2}v_x^2 + \ldots . \tag{15b}$$

Under these conditions the spatial equation (11b) is evaluated as:

$$\frac{\tilde{h}}{j}\frac{\partial \Psi_x}{\partial x} = \pm\sqrt{c^2\left(\frac{1}{w^2}-1\right)}\Psi_x = \pm\frac{v_x}{w} = \pm p_x . \tag{16}$$

So, as

$$\frac{\partial \Psi_x}{\partial x} = \pm j\frac{p_x}{\tilde{h}}\Psi_x, \text{ it follows that } \Psi_x = k_2 \exp\left[\pm\frac{p_x}{\tilde{h}}x\right]. \tag{17}$$

Equations (12) and (17) together show a complete wave function assigned to a particle moving at a constant speed. Although we have a solution, we wish to compose a wave equation which yields this solution. We note that the temporal part shows a single solution, but the spatial part shows two solutions. This result is only possible if the temporal part of the wave equation is of first order and the spatial part is of second order. It can therefore readily be verified that the wave equation that produces a solution as shown by (12) and (17) must have the following form:

$$j\tilde{h}\frac{\partial \Psi}{\partial t} + \tilde{h}^2\frac{\partial^2 \Psi}{\partial x^2} - c^2\Psi = 0 . \tag{18}$$

This equation is only slightly different from the classical form of the time dependent Schrödinger's equation for free particles. The third term in the left hand part does not occur in the classical formulation. This part however only causes an offset-frequency shift in the argument of the exponent of the temporal part. It is therefore irrelevant in the probability density function assigned to the wave function. The most striking conclusion however is that this equation is the one and only correct relativistic wave equation for free moving spinless mass particles, as all relativity laws have been fully taken into account in the derivation. It is the third term in the left hand part and the correct interpretation of the frequency of the wave function in terms of relativistic energy which makes the difference with the classical form. This equation does not show the flaws of the Klein Gordon equation ( which is of second order both in the temporal part and in the spatial part). It is therefore the author's belief that the Klein Gordon equation is mistakenly perceived as the relativistic wave equation for spinless particles. The underlying reason is the unjustified manipulation of squared momenta into second order operators on wave functions as applied in its derivation.

## Dirac's equation

Historically Dirac derived his equation for electrons in order to provide a relativistic wave equation which did not show the flaws of the Klein Gordon equation. Now we have shown that the correct intepretation of Schrödinger's equation is relativistic there does not seem any longer a justification for Dirac's mathematical manipulation that resulted in his famous and brilliant equation. So the justification for the manipulation must lie elsewhere. So, let us take another viewpoint.

Although spin of a pointmass is a meaningless concept, this is not necessarily true for a wave function assigned to a pointmass. Although there does not seem a need to suppose a spinning wavefunction for a pointmass that is only characterized by its mass, there is good reason to suppose that the wave function of a pointmass that has both mass and charge has a more complex composition. Both mass and charge are energy sources, and as a wave function is a manifestation of energy one might expect a kind of dimensionality in the wave function that represents, next to mass, the charge as well. If neverthess the wave function has to be derivable form the motion equation of a moving point-mass, next to the hypothesis of canonical of momenta into operators on wave functions, an additional hypothesis is required. So let us elaborate a bit on the idea of spinning material waves.

The most simple picture of a spinning wave function is the one with two spatially orthogonal components that are shifted temporally by 90 degees (like circular polarization of light). The spatially onedimensional equivalent of it is a wave function with two different components that are shifted in phase by 90 degrees. As not the total of the wave function is necessarily spinning, there is no particular need that the two components are equal. Let us illustrate this by the probability density function determining the probability to find the particle a some time at a certain position. This is given by:

$$Pr(x, t) = \Psi(x, t)\Psi^*(x, t) = |\Psi(x, t)|^2. \tag{18a}$$

As the temporal parts cancel, the function is positive definite, meaning that the spatial integral of the probability function is indepent of time, as it should be (at any time the particle should have to be found somewhere).

If $\Psi$ is splitted up as $\Psi_1 + \Psi_2$ just a linear split would not make any difference, but if the split is made by phase shifted components the result would be:

$$Pr(x, t) = \Psi_1\Psi^*_1 + \Psi_1\Psi^*_2 + \Psi^*_1\Psi_2 + \Psi_2\Psi^*_2 = |\Psi_1|^2 + |\Psi_2|^2.$$

The probability density function is the sum of two, and the total is positive definite. How to connect this view to the motion equation of a pointmass?

And here the brilliance of Dirac's hypothesis comes in. To keep things simple, we suppose that the resulting motion takes place under conditions of Special Relativity. This means that the momentum equation of the motion as described by (6) simplifies to:

$$p_0^2 + p_x^2 = -c^2, \tag{19}$$

or, equivalenly as: $\quad p'^2_0 + p'^2_x + 1 = 0 \quad \text{with } p'_i = \frac{p_i}{c}, i = 0, 1.$

As stated before, this expression is nothing else than Einsteins's famous energy relationship for moving particles. Dirac wrote this expression as the square of a linear relationship:

$$p'^2_0 + p'^2_x + 1 = (\alpha \cdot p' + \beta) \cdot (\alpha \cdot p' + \beta) = 0 \text{ , with } \alpha(\alpha_0, \alpha_1) \text{ and } p'(p'_0, p'_x). \tag{20}$$

thereby leaving open for the moment the numbertype of the number $\beta$ and of the components $\alpha_0$ and $\alpha_1$ of the twodimensional vector $\alpha$ .

The elaboration of the middle term is:

$$(\alpha \cdot p' + \beta) \cdot (\alpha \cdot p' + \beta) = \left(\beta + \sum_i \alpha_i p'_i\right)\left(\beta + \sum_j \alpha_j p'_j\right)$$

$$= \beta^2 + \sum_i \beta \alpha_i p'_i + \sum_j \beta \alpha_j p'_j + \sum_i \sum_j \alpha_i \alpha_j p'_i p'_j$$

$$= \beta^2 + \sum_i (\beta \alpha_i + \alpha_i \beta) p'_i + \sum_{i \neq j} (\alpha_i \alpha_j + \alpha_j \alpha_i) p'_i p'_j + \sum_i \alpha_i^2 p'^2_i. \tag{21}$$

To equate this middle term with the left hand term the following conditions should be true:

$$\alpha_i \alpha_j + \alpha_j \alpha_i = 0 \text{ if } i \neq j; \qquad \beta \alpha_i + \alpha_i \beta = 0$$

and $\qquad \alpha_i^2 = 1, \quad \beta^2 = 1 \quad \text{for } i = 0, 1. \tag{22}$

From these expressions it will be clear that the numbers $\alpha_i$ and $\beta$ have to be of special type. To this end Dirac invoked the use of the Pauli-matrices, which are defined as:

$$\sigma_1 = \begin{bmatrix} 1 & 0 \\ 0 & -1 \end{bmatrix} \qquad \sigma_2 = \begin{bmatrix} 0 & -j \\ j & 0 \end{bmatrix} \qquad \sigma_3 = \begin{bmatrix} 0 & 1 \\ 1 & 0 \end{bmatrix}. \tag{23a}$$

In addition to these also the unity matrix is required, which is defined as:

$$\sigma_0 = I = \begin{bmatrix} 1 & 0 \\ 0 & 1 \end{bmatrix} \tag{23b}$$

It can simply be verified that:

$$\sigma_1^2 = \sigma_2^2 = \sigma_3^2 = 1$$

$$\sigma_1 \sigma_2 = j\sigma_3; \; \sigma_3 \sigma_1 = j\sigma_2, \; \sigma_2 \sigma_3 = j\sigma_1 \text{ and } \sigma_i \sigma_j = -\sigma_j \sigma_i \text{ for } i \neq j. \tag{24}$$

So, squaring of the impulse relationship, as in (20), can be justified if (for instance):

$$\alpha = \alpha(\sigma_3, \sigma_1) \text{ en } \beta = \sigma_2. \tag{25}$$

Note: It may seem that the Pauli matrices can be assigned in an arbitrary order. However, this freedom does appear not to exist. The reason is that the Dirac decomposition is not the only condition that has to be fulfilled. There is also condition (18b) which states that:

$$\Psi_1 \Psi^*_2 + \Psi^*_1 \Psi_2 = 0. \tag{25a}$$

As we shall see, the assignment given by (25) will eventually yield a solution that satisfies condition (25a). In literature this assignment problem is usually overlooked and the problem is settled by quoting something as: "among the various possibilities we choose....".

As the impulse relationship is twodimensional, the wave function $\Psi$ should be twodimensional as well. Therefore $\Psi = \Psi(\Psi_1, \Psi_2)$. After transforming the impulses into operators on wave functions, the impulse relationship is transformed into the following twodimensional wave equation:

$$[\sigma_1]\begin{bmatrix}\hat{p}'_0\Psi_1\\\hat{p}'_0\Psi_2\end{bmatrix} + [\sigma_2]\begin{bmatrix}\hat{p}'_x\Psi_1\\\hat{p}'_x\Psi_2\end{bmatrix} + [\sigma_3]\begin{bmatrix}\Psi_1\\\Psi_2\end{bmatrix} = 0 ,\tag{26a}$$

or, with explicit expressions of the Pauli-matrices:

$$\begin{bmatrix}0 & 1\\1 & 0\end{bmatrix}\begin{bmatrix}\hat{p}'_0\Psi_1\\\hat{p}'_0\Psi_2\end{bmatrix} + \begin{bmatrix}1 & 0\\0 & -1\end{bmatrix}\begin{bmatrix}\hat{p}'_x\Psi_1\\\hat{p}'_x\Psi_2\end{bmatrix} + \begin{bmatrix}0 & -j\\j & 0\end{bmatrix}\begin{bmatrix}\Psi_1\\\Psi_2\end{bmatrix} = 0 .\tag{26b}$$

This reads as the following two equations:

$$\hat{p}'_0\Psi_2 + \hat{p}'_x\Psi_1 - j\Psi_2 = 0 \quad \text{and} \quad \hat{p}'_0\Psi_1 - \hat{p}'_x\Psi_1 + j\Psi_1 = 0 ,\tag{26c}$$

or, equivalently:

$$\hat{p}_0\Psi_2 + \hat{p}_x\Psi_1 - jc\Psi_2 = 0 \quad \text{and} \quad \hat{p}_0\Psi_1 - \hat{p}_x\Psi_2 + jc\Psi_1 = 0 ,\tag{26d}$$

or, in matrix terms:

$$\begin{bmatrix}\hat{p}_x & \hat{p}_0 - jc\\\hat{p}_0 + jc & -\hat{p}_x\end{bmatrix}\begin{bmatrix}\Psi_1\\\Psi_2\end{bmatrix} = 0 .\tag{26e}$$

Let us suppose that both wave functions are of the type of material waves like before in the Schrödinger-case:

$$\Psi_i(x, t) = u_i\exp\left[j\left(\beta\frac{p_x}{\tilde{h}}x - \frac{E}{\tilde{h}}t\right)\right],\tag{27}$$

wherein the constant $\beta$ may be either 1, or -1, so:

$$\beta = \pm 1.\tag{28}$$

Note that $E$ and $p_x$ are defined by (15a,b) and (16) and that they represent respectively the energy and momentum per unit of restmass in relativistic terms.

This means that the two components of the wave function have a similar behavior, but that they may differ in amplitude, and that they may be temporally phase shifted as well. The latter is the case if a phase factor is included in $u_i$, which implies that $u_i$ is supposed to be complex. So, the assumption for the wave function type matches with the requirements that a spinning wave function should satisfy.

After substitution of (27) into (26e) we find:

$$\begin{bmatrix} \beta p_x & j(E/c - c) \\ j(E/c + c) & -\beta p_x \end{bmatrix} \begin{bmatrix} u_1 \\ u_2 \end{bmatrix} = 0. \tag{29}$$

Non-trivial solutions for $u_i$ are obtained if the determinant of the matrix is zero. This is true if:

$$\frac{E^2}{c^2} = \beta^2 p_x^2 + c^2. \tag{30}$$

If we substitute in this equation the values for the relativistic energy $E$ and the relativistic momentum per unit of rest-mass as expressed by (15) and (16) we find that the condition of a zero value for the determinant reads as:

$$\beta = \pm 1, \tag{31}$$

which matches with (28). This means that the assumptions we made on the character of the wave function, as expressed by (27) and (28), are justified. This also means that a dual outcome of the determinant expression could be expected. This dual outcome simply means that the spatial part of the wave function can be a real function instead of just a complex one. And that is what we would expect from a wave function which meets the requirements for locality. Only if the wave function is spatially real it is possible to compose a spatially confined package of energy out of an ensemble of individual wave functions. So, we would have to be surprised if the outcome of the determinant expression would *not* have been dual.

This is a striking conclusion. As one knows, Dirac was surprised by a dual outcome of the determinant equation, because he interpreted it as a dual outcome for the energy. So, he was puzzled how to interprete the negative energy. In retrospect the reason for his puzzle was a misinterpretation of the energy in expression (27). If the correct real relativistic meaning had been given in agreement with a relastivistic interpretation of Schrödinger's equation, his surprise had gone soon. And let us be fair: the discovery of the positron by Anderson in 1932 can not really justify Dirac's negative outcome of energy, although scientists usually do so. Why should a mass particle which carries a positive charge instead of a negative one should have a negative motion energy instead of a positive one?

Let us now calculate the ratio of the amplitude values $u_i$. It follows now straightforwardly from (29) that

$$u_2 = \pm j \frac{v_x}{c(w + 1)} \quad \text{if} \quad u_1 = 1. \tag{32}$$

This means that the amplitude of the second component of the wave function is usually much smaller than the first component. In the non relativistic limit this component is negligeable. The imaginary value of the amplitude of the second component shows the expected phase factor. Obviously there are two possible values of the phase factor. It is therefore said that the "spin" can be positive or negative.

## Schrödinger's equation as limit case of Dirac's equation

If these views on the relativistic Schrödinger equation for spinless particles and on the relativistic Dirac's equation for particles with spin are correct it must be possible to see Schrödinger's equation as a limit case of Dirac's equation. One might expect that, by imposing a zero value on one of the two wave function components in Dirac's equation, Schrödinger's equation should follow. So, let us verify whether this is true, by the use of (26d). To avoid trivial solutions we shall suppose that $\Psi_2$ is a time independent function $K(x)$ which may or may not approach to zero. Under this condition (26d) can be written as:

$$:\frac{\tilde{h}}{j}\frac{\partial \Psi_1}{\partial x} + cK = 0 \quad \text{and} \quad \frac{\tilde{h}}{jc}\frac{\partial \Psi_1}{\partial t} - \frac{\tilde{h}}{j}\frac{\partial K}{\partial x} + jc\Psi_1 = 0. \tag{32a,b}$$

Combining (32a) and (32b) we get, after writing $\Psi_1$ as $\Psi$:

$$j\tilde{h}\frac{\partial \Psi}{\partial t} + \tilde{h}^2\frac{\partial^2 \Psi}{\partial x^2} - c^2\Psi = 0. \tag{33}$$

and this is the relativistic form of Schrödinger's equation indeed, as we derived before to (18)!

## Feynman's path integral

So far we have only considered wave equation solutions for particles moving at constant speed, i.e. wave equations for free moving particles. Mathematically it means that the components of the metric tensor in (10a,b) are equal to 1. If the particles are subject to an energy field they move in curved space-time and the metric components are no longer 1. Therefore the wave equation for particles in an energy field will be more complex. How to calculate such a wave equation? Below we wish to develop a vision on this that is based upon General Relativity.

The particles in a field move along geodesic lines, which are nothing else than straight lines in $(\xi, \tau)$-space that are transformed into so-called world lines in $(x, t)$-space. So, let us consider the motion along a small interval in $(\xi, \tau)$-space. In this small interval the particle's motion can be considered to be at constant speed, albeit that the speed depends on the coördinates of that interval. So, in that interval there is a validity of the Lorentz transform. The local validity of the Lorentz transform is one of the corner stones of General Relativity indeed. As the motion takes place along a straight line, the total motion can be splitted up in elementary motions with spatial and temporal intervals of equal size. This equal size splitting would not be possible by following the motion along equivalent elementary motions along the world-lines in $(x, t)$-space. So, it can be expected that analysis of the wave function on the basis of elementary intervals is easier in $(\xi, \tau)$-space than in $(x, t)$-space. In this case, where we have a single spatial dimension, the straight-line motion is along the $\tau$-axis as no $\xi$-distance can be covered in proper time.

Let us denote the straight line intervals by $\Delta\xi$ en $\Delta x$. If we assume a constant speed in the interval, we may state local validity of the Lorentz transform, so that:

$$\Delta\xi = \Delta\xi(x, t) = \frac{\partial \xi}{\partial x}\Delta x + \frac{\partial \xi}{\partial t}\Delta t = \frac{1}{w}\Delta x - \frac{v_x}{w}\Delta t$$

and: $$\Delta\tau = \Delta\tau(x, t) = \frac{\partial \tau}{\partial x}\Delta x + \frac{\partial \tau}{\partial t}\Delta t = -\frac{1}{w}\frac{v_x}{c^2}\Delta x + \frac{1}{w}\Delta t. \quad (34a,b)$$

Note that the Lorentz-parameter $w$ and the speed $v_x$ have a local value, so $w(x)$ and $v_x(x)$. Let us now consider the very first interval. The end points of this interval in $(\xi, \tau)$-space are $(0, \Delta\tau)$. These correspond in the $(x, t)$-space with:

$$0, \Delta\tau \to \Delta x, \Delta t = \frac{v_x}{w}\Delta\tau, \frac{\Delta\tau}{w}. \quad (35)$$

We know that the wave function of this first interval has the general format:

$$\Psi(x, t) = \Psi_t \Psi_x \;,$$

with $\quad \Psi_t = \exp\left[-j\frac{E}{\hbar}t\right]$ and $\Psi_x = \exp\left[\pm j\frac{p_x}{\hbar}x\right]$,

$$\text{wherein} \quad E = \frac{c^2}{w} \quad \text{and} \quad p_x = \frac{v_x}{w}. \quad (36)$$

We may therefore write for the wave function at the end of the first interval:

$$\Psi(0, \Delta\tau) = \exp[-j\phi_n] \;, \text{wherein} \; \phi_n = \frac{E\Delta\tau}{\hbar\,w} \pm \frac{p_x^2}{\hbar}\Delta\tau. \quad (37)$$

Apart from $p_x(x)$ and $w(x)$ the energy $E(x)$ has a local value as well.

For the wave function at the end of all $N$ intervals in succession we may write:

$$\Psi_\xi(\xi, \tau) = \Psi(0, N\Delta\tau) = \exp\left[\sum_{n=1}^{N}\exp[-j\phi_n]\right]. \quad (38)$$

For infinitesimal values of $\Delta\xi$ and $\Delta\tau$ the sum evolves into an an integral, so that:

$$\Psi_\xi(\xi, \tau) = \exp\left[-\int_0^{(\xi, \tau)} j\frac{E}{\hbar w}d\tau \pm j\frac{p_x^2}{\hbar}d\tau\right]. \quad (39)$$

We may write this wave function in terms of the parameters of the $(x, t)$-space as well. This can be done by invoking the relationships (34a,b). After substituting of these in (39) we find:

$$\Psi_\xi(\xi, \tau) \to \Psi(x, t) = \exp\left[-\int_0^{(x,t)} j\frac{E}{\hbar}dt \pm j\frac{p_x}{\hbar}dx\right]. \tag{40}$$

As noted before, $E(x)$ and $p_x(x)$ are local values. This makes the expression different from a trivial result as compared with constant values. In this latter case the expression evolves readily to the wave function of a free moving particle, as it should. Note that the integral has to be made up over a particular *x-t*-path. This path is the mapping of a straight line in $(\xi, \tau)$-space into a curve in $(x, t)$-space.

Let us now consider the phase behavior of the wave function. It clearly shows an ambiguity. This ambiguity has given rise to the antiparticle myth in quantumphysics. In fact it is not mysterious at all, if we realize ourselves that quantumtheory is a statistical theory. There is no one-to-one mapping of real particles onto individual harmonic wave functions. A real particle has to be described by an ensemble of those functions. To meet the requirements of causality and locality, each of those functions should be composed of two harmonic wave functions with opposite spatial phases. Nevertheless for ease of representation we may assign an individual wave function to an individual particle, as long as we realize ourselves that such a particle is a pseudo-particle. And if we accept such a concept of pseudo-particles we may introduce the theoretical concept of antiparticle as well. This is useful if we wish to assign a pseudo-particle to each of the two parts of an individual wave function. Each speudo-particle then has its dual antiparticle. But we should always realize ourselves that such pseudo-particles and antiparticles are nothing else than mathematical abstractions which have nothing to do with physical reality.

From (40) we conclude that the phase of the wave function along the geodesic line in $(x, t)$-space develops as:

$$\phi = -\int_0^{(x,t)} \frac{E}{\hbar}dt \pm \frac{p_x}{\hbar}dx. \tag{41}$$

The value of $E(x)$ requires some special attention. To obtain its local value, one might think that it suffices to insert in (15a) the local value for $w(x)$. This however is not correct. The reason is that the energy parameters of the successive intervals are closely interrelated and that they can not be considered to behave as if they were the energy parameters of wave equations of free moving particles. In non-relativistic mechanics this interrelationship is expressed by the concept of potential energy. In (general) relativistic mechanics this concept of potential energy does not exist. Instead there is the concept of curved space-time, as expressed by the parameters $g_{tt}$ and $g_{xx}$ of the metric tensor. The basic reason to consider the behavior of the wave function by successive inspection in elementary intervals of the geodesic curve, as we did above, is to escape from the complexity of wave equations that include $g_{tt}$ and $g_{xx}$. In fact these parameters are nothing else than the equivalent of potential energy. And somehow we have to include the influence of these parameters in our analysis. To do so, we simply account for this influence by modelling $E(x)$ such that the influence $g_{tt}$ and $g_{xx}$ is replaced by potential energy. Therefore we adapt (15a) such that it includes a potential energy $V(x)$:

$$E(x) + V(x) = \frac{c^2}{w(x)}. \tag{42}$$

By inserting this value into (41) we get:

$$\phi = -\int_0^{(x,t)} \frac{E}{\tilde{h}}dt \pm \frac{p_x}{\tilde{h}}dx = -\int_0^{(x,t)} \frac{c^2}{w\tilde{h}}dt - V(x)dt \pm \frac{p_x}{\tilde{h}}dx. \tag{43}$$

This integral can be expanded into:

$$\phi = -\int_0^{(x,t)} \frac{1}{\tilde{h}}\left(c^2 + \frac{1}{2}v^2 + \ldots\right)dt - \frac{V(x)}{\tilde{h}}dt \pm \int_0^{(x,t)} \frac{p_x}{\tilde{h}}dx. \tag{45}$$

This phase is built up by two parts: a temporal part $\phi_t$ and a spatial part $\phi_s$. The temporal part determines the time delay of the spatial package over the geodesic path. This part equals to:

$$\phi_t = -\int_0^{(x,t)} \frac{1}{\tilde{h}}\{c^2 + L(x,t)\}dt,$$

wherein: $$L(x) = \frac{1}{2}v^2(x,t) - V(x). \tag{46}$$

The function $L(x)$ is known as the *Lagrangian*. The time integral of the Lagrangian over the geodesic path is known as the *least action*. It is known and can be proven that the integral of the Lagrangian over any other path apart from the geodesic path has a larger value. Apart from the offset-phase because of $c^2$ this integral is equal to Feynman's path integral. Feynman derived this integral from a different viewpoint, thereby denoting its importance. According to his view energy-packages can travel over any path. The path with the shortest delay is the most significant one, because only packages that travel along paths very close to the one with the smallest delay add up coherently.

## On the physical appearance of electrons

Finally we wish to elaborate a bit on the wave equation of particles with spin as derived in the section on Dirac's equation. For a single spatial dimension next to the temporal dimension the end result (27,30,31) appeared to be:

$$\Psi_i(x,t) = u_i \exp\left[j\left(\beta\frac{p_x}{\tilde{h}}x - \frac{E}{\tilde{h}}t\right)\right] \text{ with } \beta = \pm 1 \text{ and } u_i = (u_1, u_2) = \left(1, \pm j\frac{v_x}{c(w+1)}\right). \tag{47}$$

So, there are two natural pairings: the spatial pairing and the spin pairing. The question is now how these wave functions match with the physical appearance of electrons and what kind of role is played by these pairings. The author wishes to view this as follows:

The electron manifests itself in two forms. One manifestation clearly is the manifestation as a real particle like any mass particle. Therefore as a spatially confined package of energy. So, this electron manifestion has to obey the laws of causality and locality. In spite of these laws it is nevertheless possible to describe this confined package of energy by a set of non-causal harmonic wave functions. The ensemble of wave functions should therefore also show the $\beta$-pairing between pairs of wave functions. It will be also clear that a real particle manifestion of the electron can not be represented by a single or a double single wave function.

This is different for the second manifestation of an electron. This second manifestation is the electron motion around the nucleus of an atom. As the spheric space is spatially confined, a single harmonic wave function is sufficient to describe the energy package of an electron. This wave function may or may not show the second pairing, i.e. the spin pairing. Both, but not more than both can represent the lowest form of energy packing of an atomic electron. In this lowest energy mode the energy of the electron spreads and fills the spherical space around the nucleus by not more than two modes of a single wave function. The Pauli exlusion principle can therefore be readily understood. The wave function can show higher modes of energy, but always in complete or incomplete pairs of two. The particle interpretation states that electrons move in discretely spaced orbits.

Now the question can be asked how it is possible that an electron bound in an atom, which is represented by a single harmonic wave function, if made free, can turn into a real particle. Or stated otherwise: how can the second manifestation can turn into the first manifestation? It is the statistical character of the theory that helps to answer the question. The answer is that there does not necessarily exist a one-to-one mapping of atomic electrons into free electrons. The theory does not say anything else than that a bunch of atomic electrons can turn into a bunch of free electrons without a one-to-one mapping, but nevertheless in such a way that the rest energy of a free electron equals the energy of an atomic electron.

### jitter of a free moving electron

As the wave function of a free electron has the additional spin component, it can be expected that the free motion of an electron is slightly different from the motion of a spinless particle. This difference can be understood by viewing the transform of a motion into a wave function in the inverse direction, i.e. by transforming a wave function into a motion. So, we start from the wave function, now written as follows:

$$\Psi(x, t) = (1 \pm j\sigma)\exp\left[j\left(\pm \frac{p_x}{\hbar}x - \frac{E}{\hbar}t\right)\right] \text{ with } \sigma = \frac{v_x}{c(w+1)}. \tag{48}$$

As any particle with mass moves with a velocity that is much smaller than velocity of light, $\sigma$ is a small value. Therefore we may apply the following approximation:

$$1 \pm j\sigma \approx \exp[\pm j\sigma]. \tag{49}$$

This enables us to write the wave function as:

$$\Psi(x, t) = \exp\left[j\left(\pm \frac{p_x}{\hbar}x - \frac{E}{\hbar}t \pm \sigma\right)\right]. \tag{50}$$

The spin manifests itself as an offset-phase in the harmonic wave function. This phase shift can be positive or negative and corresponds with a time shift of a motion of a real particle that is described by an ensemble of those functions. The spin distribution over all these wave functions has a random character and any spin can turn from negative into positive or the other way around. So, the free electron motion shows a slight amount amount of jittering, known as the "Zitterbewegung". This jittering has a "Brownian" character, and is not periodic as is sometimes incorrectly stated. To give a correct interpretation for the jitter amplitude from (50), some care has to paid. This is because of the large offset value exhibited by $E$ as a result of the rest mass energy value. By invoking (15a) we can separate the offset phase from the wave function. To this end $E$ is expanded as:

$$E = \frac{c^2}{w} = c^2\left(1 + \frac{1}{2}\frac{v_x^2}{c^2} + \ldots\right) = c^2 + \frac{1}{2}v_x^2. \quad (51)$$

Separating the offset term $c^2$, (50) is rewritten into:

$$\Psi(x, t) = \exp\left[-j\frac{c^2}{\tilde{h}}t\right]\exp\left[j\left\{\pm\frac{p_x}{\tilde{h}}x - \frac{1}{2}\frac{v_x^2}{\tilde{h}}t \pm \sigma\right\}\right]. \quad (52)$$

This can be rewritten as:

$$\Psi(x, t) = \exp\left[-j\frac{c^2}{\tilde{h}}t\right]\exp\left[j\left\{\pm\frac{p_x}{\tilde{h}}x - \frac{1}{2}\frac{v_x^2}{\tilde{h}}(t \pm \tau)\right\}\right], \text{ wherein } \tau = \frac{2\tilde{h}}{v_x^2}\sigma. \quad (53)$$

Knowing that the wave function without jitter retransforms into a motion with velocity $v_x$, the wave function with jitter transforms into a motion as:

$$x(t) = v_x(t \pm \tau) = v_x t \pm v_x \tau = v_x t \pm v_x \frac{2\tilde{h}}{v_x^2}\sigma. \quad (54)$$

Invoking (48) for $\sigma$, the jitter amplitude expressed as a position displacement appears to be:

$$v_x \frac{2\tilde{h}}{v_x^2}\sigma = \frac{2\tilde{h}}{c(w+1)} \approx \frac{\tilde{h}}{c}. \quad (55)$$

As this result applies for a particle with unit mass, this expression can be denormalized to include the electron mass $m_e$. So, the position uncertaincy $\Delta x$ for a moving electron apparently is:

$$\Delta x = \frac{\tilde{h}}{m_e c}. \quad (56)$$

This expression can be recognized from the defraction experiments of Compton. It is known as the "Compton wavelenght". The conclusion is therefore that quantumtheory predicts a position uncertaincy for a moving electron equal to the Compton wavelenght [10].

## Maxwell's wave equation

As in General Relativity photons have been modelled by Einstein as the massless limit of material particles, it can be expected that Maxwell's wave equation should somehow fit in the analysis as above. So, let us see if we may find a confirmation for this expectation.

In the section "Basics" above we learnt that the invariance of the local interval is expressed by:

$$dx_0^2 + dx_1^2 = d\tau'^2 \quad \text{wherein } x_1 = x, \, x_0 = jct \text{ and } \tau' = jc\tau. \tag{57}$$

As proper time $\tau$ for light particles is zero, this equation evaluates to:

$$\left(\frac{dx_0}{d\tau'}\right)^2 + \left(\frac{dx_1}{d\tau'}\right)^2 = 0 \text{ or, alternatively: } p'^2_0 + p'^2_x = 0. \tag{58}$$

Writing this equation according Dirac's method for dual energy particles, we get:

$$p'^2_0 + p'^2_x = (\alpha \cdot p') \cdot (\alpha \cdot p') = 0, \text{ with } \alpha(\alpha_0, \alpha_1) \text{ and } p'(p'_0, p'_x). \tag{59}$$

This imposes the following condition on $\alpha(\alpha_0, \alpha_1)$:

$$\alpha_i \alpha_j + \alpha_j \alpha_i = 0 \text{ if } i \neq j; \text{ and } \alpha_i^2 = 1, \quad \text{for } i = 0, 1. \tag{60}$$

The Pauli matrices (23a) again can be used to fulfill this condition. Making the same selection as for common mass particles we assign:

$$\alpha = \alpha(\sigma_3, \sigma_1). \tag{62}$$

So, we get from (59):

$$[\sigma_1] \begin{bmatrix} \hat{p}'_0 \Psi_1 \\ \hat{p}'_0 \Psi_2 \end{bmatrix} + [\sigma_2] \begin{bmatrix} \hat{p}'_x \Psi_1 \\ \hat{p}'_x \Psi_2 \end{bmatrix} = 0, \tag{63}$$

or, written explicitely:

$$\begin{bmatrix} 0 & 1 \\ 1 & 0 \end{bmatrix} \begin{bmatrix} \hat{p}'_0 \Psi_1 \\ \hat{p}'_0 \Psi_2 \end{bmatrix} + \begin{bmatrix} 1 & 0 \\ 0 & -1 \end{bmatrix} \begin{bmatrix} \hat{p}'_x \Psi_1 \\ \hat{p}'_x \Psi_2 \end{bmatrix} = 0. \tag{64}$$

This reads as the following two equations:

$$\hat{p}'_0 \Psi_2 + \hat{p}'_x \Psi_1 = 0 \text{ and } \hat{p}'_0 \Psi_1 - \hat{p}'_x \Psi_1 = 0, \tag{65}$$

or, equivalently:

$$:\hat{p}_0 \Psi_2 + \hat{p}_x \Psi_1 = 0 \quad \text{and} \quad \hat{p}_0 \Psi_1 - \hat{p}_x \Psi_2 = 0. \tag{66}$$

Written out as differential equations, we get

$$-\frac{1}{c}\frac{\partial \Psi_2}{\partial t} + \frac{\partial \Psi_1}{\partial x} = 0 \quad \text{and} \quad \frac{1}{c}\frac{\partial \Psi_1}{\partial t} + \frac{\partial \Psi_2}{\partial x} = 0. \tag{67}$$

The two components of the wave function have to simultaneously obey two first order partial differential equations. The question is now if we can assign a physical meaning to these two components. As we are looking for a propaga-tion equation for electromagnetic energy it is not a wild guess to suppose that these two components could represent respectively the the electrical field and the magnetic field. So, somehow the equation set (67) must have a relationship with Maxwell's rotation equations, which are formally expressed by:

$$\nabla \times \mathbf{E} + \mu_0 \frac{\partial \mathbf{H}}{\partial t} = 0 \quad \text{and} \quad \nabla \times \mathbf{H} - \varepsilon_0 \frac{\partial \mathbf{E}}{\partial t} = 0. \tag{68}$$

Maxwell's theory learns that electromagnetic wave propagation takes place as a flat wave with mutually orthogonal components for electrical field and magnetic field. We may choose the orientation of the spatial coördinate space such that the direction of the electrical field coincides with the *y*-axis, implying that the direction of the magnetic field coincides with the *x*-axis, while both are propagating along the *z*-axis. So, we have:

$$\begin{bmatrix} \mathbf{i}_x & \mathbf{i}_y & \mathbf{i}_z \\ \frac{\partial}{\partial x} & \frac{\partial}{\partial y} & \frac{\partial}{\partial z} \\ 0 & E_y & 0 \end{bmatrix} = \mathbf{i}_z \frac{\partial E_y}{\partial x} = -\mu_0 \frac{\partial \mathbf{H}}{\partial t} \quad \text{and} \quad \begin{bmatrix} \mathbf{i}_x & \mathbf{i}_y & \mathbf{i}_z \\ \frac{\partial}{\partial x} & \frac{\partial}{\partial y} & \frac{\partial}{\partial z} \\ H_x & 0 & 0 \end{bmatrix} = \mathbf{i}_z \frac{\partial H_x}{\partial x} = \varepsilon_0 \frac{\partial \mathbf{E}}{\partial t}. \tag{69}$$

So: $$\frac{\partial E_y}{\partial x} + \mu_0 \frac{\partial H_x}{\partial t} = 0 \quad \text{and} \quad \frac{\partial H_x}{\partial x} - \varepsilon_0 \frac{\partial E_y}{\partial t} = 0. \tag{70}$$

Multiplying both equations with adequate constants yields:

$$\frac{\partial E_y}{\partial x} + \frac{1}{c} \frac{\partial H_x}{\partial t} = 0 \quad \text{and} \quad \frac{\partial H_x}{\partial x} - \frac{1}{c} \frac{\partial E_y}{\partial t} = 0 \quad \text{with } c = \frac{1}{\sqrt{\varepsilon_0 \mu_0}}. \tag{71}$$

Comparing (71) and (67), we may conclude that $\Psi_1$ and $\Psi_2$ can be identified as respectively $H_x$ and $E_y$. Integrating the first equation after *t* respectively *x*, the second after *x* respectively *t* and addition yields Maxwell's wave equation for the magnetic component respectively the electric component of a free electromagnetic wave. So, as we wished to show, we may conclude that Maxwell's wave equation fits as the massless limit in the framework as developed above for mass particles [11,12]

## Conclusions

In the sections above a view has been given on quantummechanical wavefunctions as they can consistently be derived from the equity principle of Einstein's General Relativity Theory. Although considered in a single spatial dimension only and presented in simple mathematics, essentials come forward which are usually not teached in textbooks or academic courses. So, the author believes that they are overlooked, or that they even reveal historic errors. Anyhow, they helped the author to get a proper understanding of the basics of quantumwave theory. The main conlusions are:

1. The Schrödinger Equation, if properly formulated, is consistent with Relativity Theory.

2. The Klein-Gordon Equation is an historic error.

3. The Dirac Equation should not be justified by perceived shortcomings of the Schrödinger Equation and the Klein Gordon Equation, but because by including the additional energy component as represented by electric charge.

4. Spin and charge are intimately entangled. Spin is the result of charge or the other way around.

5. There is no reason to be confused about the negative energy term in Dirac's derivation of his equation.

6. Antiparticles should not be perceived as particles that exist in concrete sense. Instead they are artificial mathematical objects resulting from the pseudo-particle modelling in statistical wave mechanical theory.

7. Anderson's positron discovery should not be considered as a confirmation for the existence of antiparticles with negative energy. The positron is nothing else than a real particle with positive electric charge.

5. Feynman's path integral to enable the calculation of particles in an energy field follows readily from a general relativistic observation of the wave function. There is no need for a particle view as Feynman did.

6. The physical appearance of electrons can be understood pretty well. The view outlined above enables a comprehensable understanding of Pauli's exclusion principle, the jittering movement of free electrons and the compatibilty of the causality an locality of free real electron particles with the description of an atomic electron by a single harmonic function.

7. Maxwell's wave equation fits as the massless limit for the wave equations of mass particles with dual energy.